\documentclass[10pt,twocolumn,journal]{IEEEtran}
\usepackage{latexsym}
\usepackage{amsfonts}
\usepackage{amsbsy}
\usepackage{amsmath,amssymb}
\usepackage{times}
\usepackage{graphicx}
\usepackage{enumerate}
\usepackage[usenames]{color}
\usepackage[dvips]{pstcol}
\usepackage{epstopdf}
\usepackage{cite}
\usepackage{amsmath}
\usepackage{amssymb}
\usepackage{amsfonts}
\usepackage{graphicx}
\usepackage{epsfig}
\usepackage{psfrag}
\usepackage{color}
\usepackage{xcolor}
\usepackage{amsfonts, bm}
\usepackage{epstopdf}
\usepackage{cite}
\usepackage{color}
\usepackage{xcolor}
\usepackage{subfig}
\usepackage{algorithm}
\usepackage{stfloats}

\input epsf

\begin{document}
\title{\Large Joint User Scheduling and Resource Allocation for Millimeter Wave Systems Relying on Adaptive-Resolution ADCs}

\author{\IEEEauthorblockN{Xihan Chen, \IEEEmembership{Student Member,~IEEE},
Yunlong Cai, \IEEEmembership{Senior Member,~IEEE},\\
An Liu, \IEEEmembership{Senior Member,~IEEE}, and Lajos Hanzo, \IEEEmembership{Fellow,~IEEE},}
\thanks{
X. Chen, Y. Cai and A. Liu are with the College of Information Science and Electronic Engineering, Zhejiang University, Hangzhou 310027, China (e-mail: {chenxihan, ylcai,
anliu}@zju.edu.cn).

L. Hanzo is with the Department of Electronics and Computer Science, University of Southampton, Southampton SO17 1BJ, U.K. (e-mail: lh@ecs.soton.ac.uk).
}
}
\maketitle
\begin{abstract}
Millimeter wave (mmWave) communication systems using
adaptive-resolution analog-to-digital converters (RADCs) have recently
drawn considerable interests from the research community as benefit of
their high energy efficiency and low implementation cost.  In this
paper, we focus on the mmWave uplink using RADCs and investigate the
joint user scheduling and resource allocation problem. Specifically,
we seek to maximize the system throughput of the scheduled users by
jointly optimizing their transmit power level and hybrid combiners as
well as the number of quantization bits, subject to practical
constraints.  By relying on fractional programming (FP) techniques, we
first covert this problem into a form amenable to optimization and
exploit the specific structures in its solutions with the aid of the
so-called Ky Fan $n$-norm. Then, the resultant optimization problem is
solved using a penalty block successive concave approximation (P-BSCA)
algorithm.  Our numerical results reveal that the proposed algorithm
substantially enhances the throughput of the scheduled users compared
to the state-of-the-art benchmark schemes and provides more flexible
and efficient resource allocation control.
\end{abstract}


\IEEEpeerreviewmaketitle
\vspace{-.4cm}
\section{Introduction}
To meet the ever-increasing data rate requirements, the amalgamation
of millimeter wave (mmWave) and massive multiple-input multiple-output
(MIMO) techniques is becoming an evident trend for future wireless
networks. However, the implementation of these techniques may not be
practical, because their fully digital implementation requires a
dedicated radio frequency (RF) chain relying on power-thirsty
high-resolution analog-to-digital converters (ADCs) for each antenna
element. Hence, utilizing hybrid combiners using low-resolution ADCs
(LADCs) could be a natural technique of addressing these power consumption
concerns. Therefore substantial research efforts have been invested in
their channel estimation and beamforming design
\cite{CE1,Mo_TWC_2017,SHL_HB}.

However, the performance of mmWave systems with LADCs is limited by
the coarse quantization, especially in the low signal-to-noise ratio
(SNR) regime. To circumvent this difficulty, the authors of
\cite{RADC} first proposed a more energy-efficient mmWave receiver
architecture using adaptive-resolution ADCs (AR-ADCs) for massive MIMO
schemes. Moreover, a pair of ingenious quantization bit allocation
strategies were developed for minimizing the quantization error
effects subject to a constraint on the total ADC power. On the other
hand, user scheduling is another critical problem for mmWave systems
employing AR-ADCs, which significantly affects the interference
pattern in the uplink. Therefore, it is desirable to take them into
account when conceiving the resource allocation strategies for data
transmission, especially when the number of potential users is huge
but the available network resources are limited.  Although the authors
of \cite{LADC_Schedule} have developed a user scheduling algorithm for
mmWave systems using LADCs, there is a paucity of literature on the
joint user scheduling and resource allocation problem using
AR-ADCs. This is because the allocation of quantization bits at each
AR-ADC would exacerbate the final decision of user scheduling, thereby
substantially affecting the
communications between the base station (BS) and the users.

Motivated by these observations, we focus our attention
on the uplink of mmWave systems
using AR-ADCs and investigate the joint user scheduling and resource
allocation problem. Specifically, we seek to maximize the system
throughput of the scheduled users by jointly optimizing the transmit
power level and hybrid combiners and allocating the quantization bits,
subject to the user scheduling constraint, transmit power constraint,
the constraints related to the quantization bits, as well as the unit
modulus constraint on the elements of the analog combining matrix.

 It is quite a challenge to globally solve the problem formulated,
 which has a complex objective function (OF) relying on multiple ratio
 terms and a combinatorial constraint. By leveraging sophisticated
 fractional programming techniques \cite{FP2}, we first recast the
 original problem into an equivalent form more amenable to
 optimization. To overcome the difficulty arising from the
 combinatorial constraint, we devise a novel sparsity-enhancing
 technique for its solutions with the aid of Ky Fan $n$-norm
 \cite{KyFan1}, The benefit of this is that the choice of the
 smoothening parameters in the conventional $l_p$-norm based heuristic
 algorithms is no longer critical~\cite{LP}. Then we propose an
 efficient \emph{penalty block successive approximation} (P-BSCA)
 algorithm for the resultant problem.  Our numerical results reveal
 that the proposed algorithm achieves a remarkable performance gain
 over the relevant benchmark schemes.  \setcounter{equation}{2}
\begin{figure*}[b]
\hrulefill
\begin{equation}\label{eq:SINR}
\eta_k(\bm{\xi})=\frac{p_k|\bm{v}^H_k\mathbf{F}^H\mathbf{D}_{\rho}\mathbf{\Phi}^H\bm{h}_k|^2}{\sum_{l\neq k} p_l|\bm{v}^H_k\mathbf{F}^H\mathbf{D}_{\rho}\mathbf{\Phi}^H\bm{h}_l|^2
+|\bm{v}^H_k\mathbf{F}^H\mathbf{D}_{\rho}\mathbf{\Phi}^H|^2+\bm{v}^H_k\mathbf{F}^H\mathbf{C}_q\mathbf{F}\bm{v}_k}.
\end{equation}
\end{figure*}
\setcounter{equation}{0}
\vspace{-.3cm}
\section{System Model and Problem Formulation}
Let us now consider the uplink of a multiuser mmWave system, where $K$
single-antenna users are distributed within a specific geographical
area and the BS is equipped with a uniform linear array (ULA) of $M$
antennas and $S\ll M$ RF chains. Specifically, the BS schedules $N\leq
S$ users for transmission in each time slot, and the set of scheduled
users is denoted by $\mathcal{N}$. For convenience, we focus our
attention on the block-fading channels model, i.e.  all channels
remain time-invariant in each fading block. Then the signal received
at the BS can be expressed as
\begin{equation}
\bm{y}=\sum_{k=1}^K \sqrt{p}_k \bm{h}_k x_k+\bm{w}=\mathbf{H}\mathbf{P}^{\frac{1}{2}}\bm{x}+\bm{w},
\end{equation}
where $\mathbf{H}\triangleq[\bm{h}_1,\cdots,\bm{h}_K]$ with
$\bm{h}_k\in\mathbb{C}^M\times1$ is the uplink channel vector from
user $k$ to the BS,
$\mathbf{P}\triangleq\mathrm{diag}({p}_1,\cdots,{p}_K)$ with $p_k$ is
the transmit power level of user $k$,
$\bm{x}\triangleq[x_1,\cdots,x_K]^T$ with $x_k\sim\mathcal{CN}(0,1)$
specifying the data symbol of user $k$, and $\bm{w} \in \mathbb{C}^{M
  \times 1}$ is an additive white Gaussian noise (AWGN) vector with
zero mean and unit variance.

The BS implements a hybrid combiner to reap the full benefits of
massive MIMO and for mitigating the effects of quantization errors
imposed by the AR-ADC, which results i na reduced hardware cost and
power consumption. Let $\mathbf{\Phi}\in\mathbb{C}^{M\times S}$ denote
the BS's analog combiner, which is usually implemented using phase
shifters and its entries obey the unit modulus constraint, i.e. we
have $|\mathbf{\Phi}(m,s)|=1,\forall m,s$. Then, the signal combined
by the analog combiner can be represented by
$\bar{\bm{y}}=\mathbf{\Phi}^H\bm{y}.$ For tractability, the linear
additive quantization noise model (AQNM) is introduced for
characterizing the quantization process~\cite{AQNM}, where the
imaginary or real component of the $s$-th element in $\bar{\bm{y}}$ is
quantized by each AR-ADC using $d_s$ quantization bits. In this case,
the quantized signal can be expressed as
\begin{align}
\bm{y}_q=\mathfrak{D}(\bm{\overline{\bm{y}}})=\mathbf{D}_{\rho}\bm{\overline{\bm{y}}}+\bm{w}_q,
\end{align}
where $\mathfrak{D}(\cdot)$ refers to the element-wise quantization
operation,
$\mathbf{D}_{\rho}\triangleq\mathrm{diag}(\rho_1,\cdots,\rho_S)$ with
$\rho_s=1-\zeta_s$ is the diagonal matrix of quantization gains, and
$\zeta_s\triangleq\frac{\pi\sqrt{3}}{2}4^{-d_s}$ stands for the
normalized quantization error; $\bm{w}_q$ is the additive quantization
noise independent of $\overline{\bm{y}}$, which obeys the complex
Gaussian distribution with zero mean and covariance matrix
$\mathbf{C}_q=\mathbf{D}_{\rho}\mathbf{D}_{\zeta}\mathrm{diag}(\mathbf{\Phi}^H\mathbf{H}\mathbf{P}\mathbf{H}^H\mathbf{\Phi}+\mathbf{\Phi}^H\mathbf{\Phi})
$, where
$\mathbf{D}_{\zeta}\triangleq\mathrm{diag}(\zeta_1,\cdots,\zeta_S)$. Note
that in contrast to the conventional fixed-resolution ADCs, AR-ADCs
can adapt the number of quantization bits to the propagation
characteristics, thereby providing additional flexibility for these
mmWave systems.

Then, the quantized signal $\bm{y}_q$ is successively processed by the
BS's digital combiner $\mathbf{F}\in\mathbb{C}^{S\times S}$ and the
linear receiver beamformer $\bm{v}_k\in\mathbb{C}^S$ so as to mitigate
the multiuser interference and alleviate the quantization loss, which
yields the recovered signal of user $k$ in the form of:
\begin{equation}
\hat{x}_k=\bm{v}^H_k\mathbf{F}^H\mathbf{D}_{\rho}\mathbf{\Phi}^H\mathbf{H}\mathbf{P}^{\frac{1}{2}}\bm{x}_k+\bm{v}^H_k\mathbf{F}^H\mathbf{D}_{\rho}\mathbf{\Phi}^H \bm{w}
+\bm{v}^H_k\mathbf{F}^H\bm{w}_q.\nonumber
\end{equation}
For ease of exposition, we let $\bm{\phi}$ denote the phase vector of
the BS's vectorized analog combiner $\mathrm{vec}(\mathbf{\Phi})$,
$\bm{p}\triangleq[p_1,\cdots,p_K]^T$,
$\bm{d}\triangleq[d_1,\cdots,d_S]^T$,
$\bm{v}\triangleq[\bm{v}^T_1,\cdots,\bm{v}^T_K]^T$,
$\bm{f}=\mathrm{vec}(\mathbf{F})$, and
$\bm{\xi}\triangleq[\bm{p}^T,\bm{\phi}^T,\bm{d}^T,\bm{v}^T,\bm{f}^T]^T$. As
such, the achievable throughput of user $k$ can be expressed as
$r_k(\bm{\xi})=\log_2\left(1+\eta_k(\bm{\xi})\right)$, where
$\eta_k(\bm{\xi})$ stands for the signal-to-interference-noise ratio
(SINR) of user $k$ defined in \eqref{eq:SINR} at the bottom of this
page.  \setcounter{equation}{4}
\begin{figure*}[b]
\hrulefill
\begin{align}
 f_k(\bm{\xi},\bm{\eta},\bm{\nu})&=2\mathfrak{R}\{\sqrt{p_k(1+\eta_k)}\nu^H_k\bm{v}^H_k\mathbf{F}^H\mathbf{D}_{\rho}\mathbf{\Phi}^H\bm{h}_k\}+\log_2(1+\eta_k)-\eta_k
 -\nu^H_k\nu_k \omega_k(\bm{\xi}),
 \label{eq:objF1}\\
 \omega_k(\bm{\xi})&=\sum_{l=1} p_l|\bm{v}^H_k\mathbf{F}^H\mathbf{D}_{\rho}\mathbf{\Phi}^H\bm{h}_l|^2
+|\bm{v}^H_k\mathbf{F}^H\mathbf{D}_{\rho}\mathbf{\Phi}^H|^2+\bm{v}^H_k\mathbf{F}^H\mathbf{C}_q\mathbf{F}\bm{v}_k.
  \label{eq:objF2}
\end{align}
\end{figure*}
\setcounter{equation}{3}

 The key observation is that the uplink interference pattern heavily
 relies on the choice of which specific users to schedule in this time
 slot. To this end, we seek to maximize the system throughput among
 the scheduled users by jointly optimizing the uplink scheduling,
 hybrid combiner design, and quantization bits allocation under
 practical constraints. Instead of directly determining the discrete
 scheduling variables, we treat the uplink users in an implicit manner
 with the aid of power control, depending on whether the
 transmit power level of user $k$ is positive.
 This optimization problem is formulated as:
\begin{subequations}\label{eq:oriP}
\begin{align}
\mathcal{P}: \mathop{\max}_{\bm{\xi}}&\quad \sum_{k=1}^K r_k(\bm{\xi}) \label{eq:oriP_obj}\\
\textrm{s.t.}
& \quad \|\bm{p}\|_0=N,\label{eq:numSU}\\
& \quad 0 \leq p_k \leq P_k^{\mathrm{max}}, \forall k,\label{eq:powerBudget}\\
& \quad  d_s^{\mathrm{min}} \leq d_s \leq d_s^{\mathrm{max}} \text{\ is an integer}, \forall
	s,\label{eq:d_limit_1}\\
& \quad \sum_{s=1}^S d_s\leq S{d}_{\mathrm{avg}}
\label{eq:d_limit_2},\\
& \quad \bm{\phi} \in \Upsilon \triangleq  [0,2\pi]^{MS},\label{eq:phaseC}
\end{align}
\end{subequations}
where $\|\cdot\|_0$ refers to the $l_0$-norm operator,
$P_k^{\mathrm{max}}$ specifies the maximum transmit power of user $k$,
$d_s^{\mathrm{min}}$ and $d_s^{\mathrm{max}}$ respectively denote the
minimum and maximum number of quantization bits in the RF chain $s$,
and ${d}_{\mathrm{avg}}$ is the average number of quantization bits
across the different RF chains. The constraint~\eqref{eq:numSU}
ensures that the number of users scheduled in each time slot is
equivalent to $N$, while the constraint~\eqref{eq:d_limit_2}
represents the quantization bits budget at the BS. Finally, the
constraint~\eqref{eq:phaseC} corresponds to the unit modulus
constraint on the elements of the analog combining matrix.
\vspace{-.2cm}
\section{Penalty Block Successive Concave Approximation Algorithm}
Problem $\mathcal{P}$ is challenging to handle because 1) the
optimization variables $\bm{\xi}$ are intricately coupled in the
non-concave OF of~\eqref{eq:oriP_obj}; 2) the discrete
variables involved in the quantization bits allocation and the presence of
$l_0$-norm in~\eqref{eq:numSU} make the feasible set non-concave,
which further complicates its solution.  In this section, we propose a
novel P-BSCA algorithm which efficiently combines the penalty method
with the block successive concave approximation (BSCA) method~\cite{SCA}
to find the stationary solution of problem $\mathcal{P}$.

\subsection{Problem Reformulation}
To facilitate an efficient algorithmic design, we first exploit the complex
quadratic and Lagrangian dual transform~\cite{FP2} for recasting problem
$\mathcal{P}$ into a series of simple equivalent problems:
\begin{align}\label{eq:P1}
\mathcal{P}_1: \mathop{\max}_{\bm{\xi},\bm{\eta},\bm{\nu}}\quad \sum_{k=1}^K f_k(\bm{\xi},\bm{\eta},\bm{\nu})\quad
\textrm{s.t.} \quad \eqref{eq:numSU}-\eqref{eq:phaseC},\nonumber
\end{align}
where the objective function $ f_k(\bm{\xi},\bm{\eta},\bm{\nu})$
associated with $\omega_k(\bm{\xi})$ is defined in
\eqref{eq:objF1}-\eqref{eq:objF2} at the bottom of the next page,
$\bm{\eta}=[\eta_1,\cdots,\eta_K]^T$ stands for the auxiliary variable
introduced for the SINR within the rate expression, and
$\bm{\nu}=[\nu_1,\cdots,\nu_K]^T$ represents the auxiliary variables
introduced to achieve the desired decoupling between the
numerator and denominator in the SINR.\setcounter{equation}{6}

To make the problem tractable, we relax the discrete constraint on the
number of quantization bits $d_s$ into a continuous one, yielding
\begin{equation}\label{eq:d_limit_3}
d_s^{\mathrm{min}} \leq d_s \leq d_s^{\mathrm{max}},
\end{equation}
and then round the solution $d^{\star}_s$ of the relaxed problem to
its nearest integer \cite{THCF} as follows 
	\begin{equation}
	\bar{d}_s(\varepsilon)=\left\{
	\begin{aligned}
	\overset{} \lfloor{d_s^{\star}}\rfloor&,
	~~~~~\text{if}~d_s^{\star}-\lfloor{d_s^{\star}}\rfloor \leq \varepsilon  \\
	\lceil{d_s^{\star}}\rceil&,~~~~~\text{otherwise,}\\
	\end{aligned}
	\right.\forall s,
	\end{equation}
where the hyper-parameter $\varepsilon\in[0,1]$ is efficiently
searched via the bisection method, so that we have $\sum_{s=1}^S
\bar{d}_s(\varepsilon)\leq S{d}_{\mathrm{avg}}$.

It is worth noting that solving problem $\mathcal{P}_1$ is still
difficult due to the non-concave $l_0$-norm
constraint~\eqref{eq:numSU}. A promising solution is to leverage the
smoothened $l_p$-norm~\cite{LP} followed by the iteratively reweighted
$l_2$-norm minimization algorithm to construct a tight surrogate
function for the $l_0$-norm, thereby inducing the sparsity structure
in uplink power control. However, the convergence properties of the
$l_p$-norm based algorithms are crucially dependent on the specific
choice of the smoothening parameters, which may not be suitable for
practical implementations due to the associated dynamic system
requirements. To this end, we devise a novel technique of enhancing
the sparsity structures in the solutions for problem $\mathcal{P}_1$
with the aid of the Ky Fan $n$-norm
of~\cite{KyFan1,KyFan2}. Specifically, we represent the $l_0$-norm in
form of the difference between the $l_1$-norm and Ky Fan $n$-norm as
follows:
\begin{equation}
\|\bm{p}\|_0=\min\{n:\|\bm{p}\|_1-\|\bm{p}\|_n=0,\forall 0\leq n \leq N\},
\end{equation}
where $\|\bm{p}\|_1\triangleq\sum_{k=1}^K|p_k|$ stands for the $l_1$-norm, and $\|\bm{p}\|_n$ represents the Ky Fan $n$-norm given by the sum of largest $n$ absolute values, i.e.,
\begin{equation}
\|\bm{p}\|_n=\sum_{i=1}^n |p_{\chi(i)}|,
\end{equation}
where $\chi$ specifies the permutation of $\{1,\cdots,K\}$ in
descending order such that we have
$|p_{\chi(1)}|\geq\cdots\geq|p_{\chi(K)}|$. Using the above notations, the
constraint \eqref{eq:numSU} reduces to $\|\bm{p}\|_1-\|\bm{p}\|_N=0$.

Now, we are ready to apply the penalty method to solve problem
$\mathcal{P}_1$. For the problem at hand, the first step is to
incorporate the penalty term associated with the equality constraint
$\|\bm{p}\|_1=\|\bm{p}\|_N$ and to obtain the penalized version of
$\mathcal{P}_1$ as
\begin{subequations}\label{eq:P2}
\begin{align}
\mathcal{P}_2: \mathop{\max}_{\bm{\xi},\bm{\eta},\bm{\nu}}&\quad \sum_{k=1}^K f_k(\bm{\xi},\bm{\eta},\bm{\nu})-\lambda(\|\bm{p}\|_1-\|\bm{p}\|_N)\label{eq:P2Obj}\\
\textrm{s.t.}& \quad \eqref{eq:powerBudget}-\eqref{eq:phaseC},
\end{align}
\end{subequations}
where $\lambda$ is a positive penalty parameter that characterizes the
cost of violating the equality constraint.

\begin{algorithm}[t]
		\caption{\label{alg:PBSCA}Proposed P-BSCA Algorithm for Problem \eqref{eq:oriP}}
		
		\textbf{Initialization:}{\small{} Initialize the algorithm with a feasible point $\bm{\xi}^0$.  Set the accuracy tolerance $\beta$, the maximum inner iteration
			number $I$, the maximum outer iteration
			number $T$, $t=0$, $i=0$, $\alpha>1$, and $\lambda^0>0$.}{\small\par}
		
		\textbf{Repeat}
		
		\textbf{~~Repeat}
		
		\textbf{\small{}\,\,\,\,\,\,\,\,\,\,-~}{\small{}Update
		$\bm{\eta}^{i+1}$, $\bm{\nu}^{i+1}$, and $\bm{f}^{i+1}$ by applying its first-order}

        \textbf{\small{}\,\,\,\,\,\,\,\,\,\,~~}{\small{}optimality condition in sequence}.

		\textbf{\small{}\,\,\,\,\,\,\,\,\,\,-~}{\small{}Construct a surrogate function $\hat{g}^i(\bm{d})$ and update $\bm{d}^{i+1}$ }

        \textbf{\small{}\,\,\,\,\,\,\,\,\,\,~~}{\small{}by solving problem \eqref{eq:Pd}}.

		\textbf{\small{}\,\,\,\,\,\,\,\,\,\,-~}{\small{}Construct a surrogate function $\hat{g}^i(\bm{\phi})$ and update $\bm{\phi}^{i+1}$ }

        \textbf{\small{}\,\,\,\,\,\,\,\,\,\,~~}{\small{}according to \eqref{eq:updatePhi}}.

		\textbf{\small{}\,\,\,\,\,\,\,\,\,\,-~}{\small{}Update $\bm{p}^{i+1}$ by solving problem \eqref{eq:Pp}}.

		\textbf{~~Until }{\small{}the value of \eqref{eq:P2Obj} converges or reaching the maximum }

		\textbf{~~~~~~~~~}{\small{}inner iteration number $I$. Otherwise, let $i\leftarrow i+1$}.

		\textbf{~~Update }{\small{}the penalty parameter: $\lambda^{t+1}=\alpha\lambda^t$}.
		
		\textbf{Until }{\small{} the value of the penalty term is less than $\beta$ or $t\geq T$}.

		\textbf{~~~~~~~~}{\small{}Otherwise, let $t\leftarrow t+1$}.
	
	\end{algorithm}
\subsection{The Algorithm Proposed for Solving Problem $\mathcal{P}_2$}
In this subsection, we present the proposed P-BSCA algorithm for solving
problem $\mathcal{P}_2$, mainly consisting of twin
loops. Specifically, we increase the value of the penalty parameter
$\lambda$ to reduce the equality constraint violation at each outer
iteration, while the BSCA method is utilized for updating the
optimization variables in different blocks within the inner
iteration. Note that the constraints in problem $\mathcal{P}_2$ are
separable, consequently we can partition the design variables into six
independent blocks. Hereinafter, we introduce the superscripts $i$ and $t$
for representing the variables associated with the $i$-th
inner and the $t$-th outer iteration, respectively. Then we elaborate on the
implementation details of the proposed P-BSCA algorithm at the $i$-th
inner iteration within the $t$-th outer iteration.

 \emph{1) \textbf{Optimization of} $\bm{\eta}$}: When fixing the other
 variables, the optimal $\bm{\eta}^{\star}$ can be uniquely determined
 by examining the first-order optimality condition, which is defined
 in~\eqref{eq:SINR}.

\emph{2) \textbf{Optimization of} $\bm{\nu}$}: The subproblem
w.r.t. $\bm{\nu}$ can be further decoupled on a per-user basis, and
each one is an unconstrained quadratic optimization problem, which can
be efficiently solved by setting $\partial f_k/\partial \nu_k=0$, that
is
  \begin{equation}
  \nu^{\star}_k={\omega^{-1}_k(\bm{\xi})}\sqrt{p_k(1+\eta_k)}\bm{v}^H_k\mathbf{F}^H\mathbf{D}_{\rho}\mathbf{\Phi}^H\bm{h}_k.
  \end{equation}

\emph{3) \textbf{Optimization of} $\bm{f}$}: Similar to the subproblem
w.r.t. $\bm{\nu}$, the subproblem w.r.t. $\bm{f}$ is also
unconstrained as well as quadratic, and can be solved by checking its
first-order optimality condition. The details are omitted due to the
limited space.

 \emph{4) \textbf{Optimization of} $\bm{d}$}: Now we turn attention to
 the subproblem w.r.t. $\bm{d}$, which features a non-concave OF
 subject to the linearly coupled constraint~\eqref{eq:d_limit_2}. To
 handle this complex problem, we resort to the SCA method for
 approximating the OF as a sequence of concave surrogate
 functions, which is given by
 \begin{align}
 \hat{g}^i(\bm{d})=\sum_{k=1}^K \left(f_k(\bm{d}^{i})+ \nabla^T_{\bm{d}}f_k(\bm{d}^i)(\bm{d}-\bm{d}^{i})\right)-\tau_d\|\bm{d}-\bm{d}^{i}\|^2,\nonumber
 \end{align}
where $\nabla_{\bm{d}}f_k(\bm{d}^i)$ is the partial derivative of
$f_k(\bm{d})$ w.r.t. $\bm{d}$ at the current point $\bm{d}^{i}$, and
$\tau_d$ is a positive constant. Note that the term
$\tau_d\|\bm{d}-\bm{d}^{i}\|^2$ is added to ensure that $
\hat{g}^i_k(\bm{d})$ is a lower bound of the original objective
function, which plays a crucial role in guaranteeing the algorithm's
convergence. Thus, finding the optimal $\bm{d}^{\star}$ amounts to
solving the following linearly constrained quadratic problem:
\begin{align}\label{eq:Pd}
\mathop{\max}_{\bm{d}}\quad  \hat{g}^i(\bm{d})\quad
\textrm{s.t.} \quad \eqref{eq:d_limit_2}~\mathrm{and}~\eqref{eq:d_limit_3},
\end{align}
which can be efficiently solved by the generic interior-point method
using off-the-shelf solvers, such as CVX.

 \emph{5) \textbf{Optimization of} $\bm{\phi}$}: Let us now consider
 the subproblem w.r.t. $\bm{\phi}$, which can be formulated as
 $\mathop{\max}_{\bm{\phi}\in\mathbf{\Upsilon}}\quad \sum_{k=1}^K
 f_k(\bm{\phi})$. Following the same approach as used for updating
 $\bm{d}$, a concave surrogate function $ \hat{g}^i(\bm{\phi})$ is
 judiciously constructed to circumvent the difficulty arising from the
 non-concave nature of the OF, which can be expressed as
 \begin{align}
 \hat{g}^i(\bm{\phi})=\sum_{k=1}^K \left(f_k(\bm{\phi}^{i})+ \nabla^T_{\bm{\phi}}f_k(\bm{\phi}^i)(\bm{\phi}-\bm{\phi}^{i})\right)-\tau_{\phi}\|\bm{\phi}-\bm{\phi}^{i}\|^2,\nonumber
 \end{align}
 where $\tau_{\phi}$ is a positive constant and
 $\nabla_{\bm{\phi}}f_k(\bm{\phi}^i)$ is the partial derivative of
 $f_k(\bm{\phi})$ w.r.t. $\bm{\phi}$ at the current point
 $\bm{\phi}^{i}$.

 Consequently, the optimal solution $\bm{\phi}^{\star}$ of the
 approximated problem
 $\mathop{\max}_{\bm{\phi}\in\mathbf{\Upsilon}}\quad
 \hat{g}^i(\bm{\phi})$ is equivalent to the projection of the gradient
 $\bm{\phi}^{i}+{ \nabla^T_{\bm{\phi}}f_k(\bm{\phi}^i)}/{\tau_{\phi}}$
 onto the box feasible region $\mathbf{\Upsilon}$, which yields a
 closed-form solution as
 \begin{equation}\label{eq:updatePhi}
 \bm{\phi}^{\star}=\mathbb{P}_{\mathbf{\Upsilon}}[\bm{\phi}^{i}+{ \nabla^T_{\bm{\phi}}f_k(\bm{\phi}^i)}/{\tau_{\phi}}],
 \end{equation}
 where $\mathbb{P}_{\mathbf{\Upsilon}}[\cdot]$ refers to the projection over the feasible set $\mathbf{\Upsilon}$.

 \emph{6) \textbf{Optimization of} $\bm{p}$}: To fully exploit the
 intrinsic structure of the subproblem w.r.t. $\bm{p}$, we first
 rewrite its non-concave OF into a
 difference-of-concave (DC) form, i.e.,
\begin{align}
\sum_{k=1}^K f_k(\bm{p})-\lambda(\|\bm{p}\|_1-\|\bm{p}\|_N)=g(\bm{p})-h(\bm{p}),
\end{align}
where $g(\bm{p})$ and $h(\bm{p})$ respectively are strongly concave
functions given by
\begin{align}
g(\bm{p})&=\sum_{k=1}^K f_k(\bm{p})-\lambda\|\bm{p}\|_1-\tau_p \|\bm{p}\|^2,\\
h(\bm{p})&=-\lambda\|\bm{p}\|_N-\tau_p \|\bm{p}\|^2.
\end{align}
By linearizing the strongly concave function $h(\bm{p})$ based on the
first-order Taylor expansion and the current point $\bm{p}^i$, we can
obtain
\begin{align}
\hat{h}^i(\bm{p})=h(\bm{p}^i)+\partial^T_{\bm{p}}h(\bm{p}^i)(\bm{p}-\bm{p}^i),
\end{align}
where $\partial_{\bm{p}}h(\bm{p}^i)$ stands for the subgradient of
$h(\bm{p})$ w.r.t. $\bm{p}$ at the current point
$\bm{p}^i$. Specifically, the subgradient of $h(\bm{p})$
w.r.t. $\bm{p}$ can be analytically computed as
\begin{equation}
\partial_{\bm{p}}h(\bm{p})=-\tau_p \bm{p}-\lambda\partial\|\bm{p}\|_N,
\end{equation}
where $\partial\|\bm{p}\|_N$ is the subgradient of $\|\bm{p}\|_N$ calculated as

$j$-th entry of  	$\partial\|\bm{p}\|_N=\left\{
	\begin{aligned}
	&\text{sgn}(p_j),
	~\text{if}~|p_j|\geq|p_{\chi(N)}|  \\
	&0,~~~~~~~~\text{otherwise.}\\
	\end{aligned}
	\right.$

Hence, in the $i$-th iteration of the proposed P-BSCA algorithm, we
have the following approximated convex problem:
\begin{align}\label{eq:Pp}
\mathop{\max}_{\bm{p}}\quad  g(\bm{p})-\hat{h}^i(\bm{p})\quad
\textrm{s.t.} \quad \eqref{eq:powerBudget},
\end{align}
which can be uniquely determined by standard convex optimization methods.

\subsection{Complete Algorithm}
According to the above derivations, we summarize the proposed P-BSCA
procedure in Algorithm~\ref{alg:PBSCA}. Here, we remark that by
appropriately tuning the penalty parameter in each outer iteration,
the limiting point $\bm{\xi}^{\star}$ generated by the proposed P-BSCA
algorithm would essentially meet the equality
constraint~\eqref{eq:numSU}. As such, we can show that the proposed
P-BSCA algorithm converges to a stationary solution of problem
$\mathcal{P}$. The proof is similar to that of \cite{P-CCCP}, and we
hence omit the details for simplicity.  Let us now further analyze the
computational complexity. In each iteration of the proposed P-BSCA
algorithm, we solve the subproblems for the six blocks of variables
sequentially.  Then, the overall computational complexity of the
proposed algorithm is dominated by updating $\{\bm{f},\bm{\phi}\}$ and
is given by $O(T_1I_1(S^6+S^2M+KM^2))$, where $I_1$ and $T_1$
respectively denote the maximum inner and outer iteration numbers.

\vspace{-.3cm}
\section{Numerical Results}
This section presents our numerical results for quantifying the performance
of the proposed algorithm, whilst providing essential insights. For all
simulations, unless otherwise specified, we consider a single-cell
network configuration of radius $r=500$ m, where a total of $K=40$
candidate users are randomly distributed and the BS is
equipped with $M=96$ antennas and $S=32$ RF chains to schedule $N=16$
users for their uplink transmission. We adopt the extended Saleh-Valenzuela
geometric model for our mmWave channels~\cite{mmWchannel}, where the
large-scale path loss of the user $k$-BS link is given by
$\gamma_k[\mathrm{dB}]=72+29.2\log_{10}\mu_k+\psi$, $\mu_k$ stands for
the corresponding distance, and $\psi\sim\mathcal{CN}(0,1)$ is the
log-normal shadowing. The channel bandwidth is $10$ MHz, and the
background noise is $-174$ dBm/Hz. Furthermore, we set
$P^{\mathrm{max}}_k=10$ dBm, $d^{\mathrm{max}}_s=8$,
$d^{\mathrm{min}}_s=1$, and $d_{\mathrm{avg}}=3$ \cite{SHL_WCL}. For
the proposed P-BSCA algorithm, we choose $\lambda^0=10^{-3}$ and
$\alpha=1.8$. Three schemes are included as benchmarks: 1) the smooth
$l_p$/$l_2$ approximation (SA) scheme of~\cite{KyFan2}, which adopts the
quadratic form of the weighted mixed $l_p$/$l_2$-norm for inducing the
sparsity in user scheduling and jointly optimizes the other variables
by maximizing the system throughput.  2) the  uniform allocation (UA)
scheme of~\cite{LADC_Schedule}, where the LADCs with uniform quantization
bits are implemented and all variables are jointly optimized for
maximizing the system's throughput.
3) random scheduling (RS) scheme, where the scheduled users are
randomly selected and the other variables are optimized for maximizing
the system throughput.

\begin{figure}[t]
		\centering
		\includegraphics[width=8.5cm]{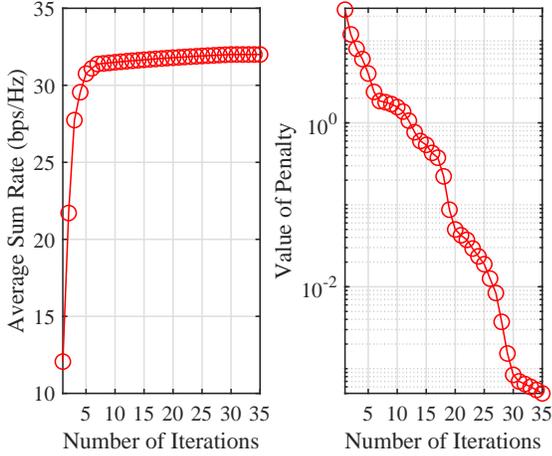}
		\caption{Convergence performance of the proposed P-BSCA algorithm: (a) average sum rate; (b) the value of penalty.
		}\label{fig:converge}
	\end{figure}

We commence by examining the convergence behavior of the proposed
P-BSCA algorithm. Fig.~\ref{fig:converge} (a) and (b) respectively
plot an instance of the average sum rate and the value of penalty
versus the number of iterations. It is observed that the proposed
P-BSCA promptly converges to a stationary solution within a few
iterations, while the penalty value reduces below a threshold of
$10^{-3}$ at the same time. These results demonstrate the ability of
the proposed P-BSCA algorithm to efficiently handle the combinatorial
constraint~\eqref{eq:numSU} of problem $\mathcal{P}$.

\begin{figure}[t]
\centering
\includegraphics[width=8.5cm]{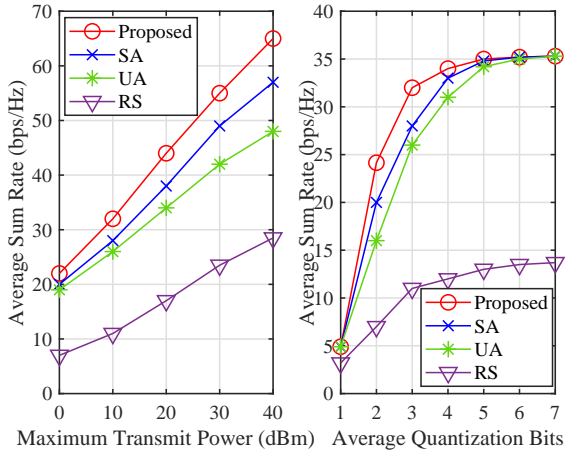}
\caption{Average sum rate versus (a)  maximum transmit power $P^{\mathrm{max}}$ and (b) the number of average quantization bits $d_{\mathrm{avg}}$. }
\label{fig:compare}
\end{figure}

Fig. \ref{fig:compare} (a) depicts the average sum rate versus the
maximum transmit power $P^{\mathrm{max}}$ for different schemes. We
notice that the average sum rate achieved by all schemes is
monotonically increasing with the maximum transmit power.
Furthermore, it is observed that the proposed P-BSCA scheme can
outperform all the other competing schemes, and its improvement becomes
more significant as the maximum transmit power increases.
The reason for this trend is that the proposed P-BSCA scheme can
exploit the distinctive channel characteristics of different candidate
users to facilitate more efficient user scheduling and resource allocation
strategies, which also demonstrates the importance of our joint
optimization based design. In a nutshell, it appears that for mmWave
systems having ample transmit power, our proposed P-BSCA scheme
is particularly attractive from an optimum resource allocation
perspective.

In Fig. \ref{fig:compare} (b), we plot the average sum rate versus the
number of average quantization bits $d_{\mathrm{avg}}$ for different
schemes. It is interesting to note that the average sum rates of the
proposed P-BSCA, SA, and UA schemes nearly coincide both at low and
high average number of quantization bits. This is due to the following
reasons: 1) For a low average number of quantization bits there is no
additional freedom for adapting the allocation of quantization bits
depending on the specific propagation conditions of the different
users. 2) When the number of average quantization bits is sufficiently
high, the quantization errors due to ADCs can be neglected and thus it
is no longer the primary bottleneck of mmWave systems. Additionally,
we observe that the proposed P-BSCA scheme generally attains a higher
average sum rate than the SA/UA scheme for a moderate average number
of quantization bits, thanks to our novel sparsity enhancement
approach based on the Ky Fan $n$-norm and to the more flexible
quantization bit allocation. On the other hand, it can be seen that
the performance of the RS scheme is much worse than that of all the
other competing schemes, due to the lack of a more efficient user
scheduling strategy.
\vspace{-.25cm}
\section{Conclusion}
In this contribution, we have investigated the joint user scheduling
and resource allocation problem of mmWave systems using
AR-ADCs. Specifically, we maximized the system throughput of the
scheduled users by jointly optimizing the transmit power level and
hybrid combiners as well as allocating the quantization bits under
some practical constraints.
optimized to maximize the sum throughput of the scheduled users under
some practical constraints.  To solve such a non-convex combinational
problem efficiently, we conceived a novel P-BSCA iterative algorithm.
Finally, our numerical results demonstrate the efficiency of the proposed algorithm.

\end{document}